\documentstyle[prb,aps,epsf]{revtex}
\include{epsf}
\newcommand{\LSCO}{{La$_{2-x}$Sr$_{x}$CuO$_{4}$}}%
\begin{document}
\draft
\preprint{Draft 1.0}
\twocolumn[\hsize\textwidth\columnwidth\hsize\csname @twocolumnfalse\endcsname
%
%
\title{Role of two-dimensional electronic state in superconductivity 
       in \LSCO}
\author{Fumihiko Nakamura, Tatsuo Goko, Junya Hori, Yoshinori Uno, 
        Naoki Kikugawa, and Toshizo Fujita}
\address{Department of Quantum Matter, ADSM, Hiroshima University,
         Higashi-Hiroshima 739-8526, Japan}
\date{\today}
\maketitle
%
%
\begin{abstract}
\hspace*{1.5mm}
We have measured out-of-plane resistivity $\rho_c$ for \LSCO{} under anisotropic pressure. 
$c$-axis compression, which decreases $\rho_c$, reduces $T_{\rm c}$ drastically, whereas $c$-axis extention, which increases $\rho_c$, enhances $T_{\rm c}$ from 38K at ambient pressure to 51.6K at 8GPa. 
We find that the variation of $T_{\rm c}$ scales as a function of $\rho_c$, and that the $c$-axis pressure coefficient is much stronger than the $ab$-axis one. 
These imply that $T_{\rm c}$ depends primarily on the interlayer, rather than the in-plane, lattice parameter.
\end{abstract}
%
%
%
%
\pacs{PACS numbers: 74.62.Fj, 74.25.-q, 74.72.Dn, and 74.62.-c }]
\narrowtext

Although a large number of experimental and theoretical investigations indicate that strong two dimensionality of the normal state is one of key factors in high-$T_{\rm c}$ superconductivity, \cite{P.W.Anderson_PRL_88,P.B.Littlewood_PRB_92,%
H.Fukuyama_JLTP_94} 
it remains to be critically examined whether the cuprates are essentially two-dimensional (2D) metals or three-dimensional (3D) metals with strong anisotropy. 
Strong two-dimensionality in the cuprates is after expressed in exotic behavior of the out-of-plane resistivity, $\rho_{c}$, {\it e.g.} a low-$T$ upturn in $\rho_{c}(T)$ for under-doped high-$T_{\rm c}$ cuprates. 
\cite{N.P.Ong_PC_94,%
Y.Ando_PRL_95,%
N.E.Hussey_PRB_98,F.Nakamura_PRB_96}
Interlayer coupling, as reflected in $\rho_{c}$, is strongly related to the 2D-electronic state.

Low dimensionality is generally known as a destructive factor for long-range order such as superconductivity. 
Moreover, a theoretical model for high-$T_{\rm c}$ cuprates \cite{J.M.Wheatly_PRB_88} indicates that $T_{\rm c}$ enhancement is caused not only by strong in-plane correlation but also by increase of interlayer coupling. 
This suggests that a strongly 2D electronic state, {\it i.e.} weak interlayer coupling, 
should weaken high-$T_{\rm c}$ superconductivity. 
In this paper, we explore how the two-dimensionality in the electronic state is reflected in $T_{\rm c}$.

Pressure, $P$, experiments on high-$T_{\rm c}$ cuprates have received much attention as a technique which can control the two-dimensionality or the lattice parameters which govern $T_{\rm c}$. 
In many high-$T_{\rm c}$ cuprates, however, it is difficult to clarify the effect of the interlayer hopping transfer $t_c$ on $T_{\rm c}$, 
because $T_{\rm c}$ is usually governed by a change of carrier concentration caused by applying $P$.\cite{H.Takahashi_SHS_96}
In contrast, \LSCO{} (LSCO) is a suitable system to investigate this question 
because the carrier number depends only weakly on $P$.
\cite{C.Murayama_PC_91,G.Q.Zheng_PC_95}
In LSCO, indeed, the $P$ dependence of $\rho_{c}$, which is much stronger than that of the in-plane resistivity $\rho_{ab}$, 
is interpreted in terms of a change of interlayer distance.
\cite{F.Nakamura_PRB_96}
In this work, we have investigated the $P$ dependence of $\rho_{c}$ for LSCO 
in order to understand the role of the two dimensionality in the superconductivity.

The presence of a structural transition from tetragonal to orthorhombic at $T_{\rm d}$ prevents us from observing the generic behavior of $\rho_{c}(T)$ and $T_{\rm c}$ in LSCO with $x < 0.2$. 
For instance, a change of slope is clearly observed in the $\rho_{c}(T)$ at $T_{\rm d}$.
\cite{Y.Nakamura_PRB_93}
Moreover, $T_{\rm c}$ in LSCO is suppressed by the orthorhombicity. \cite{Goko_JPS} 
It has been reported that $T_{\rm d}$ in LSCO can be suppressed by applying high $P$. \cite{S.Nakayama_PC_94,H.J.Kim_PC_88}
Therefore, we have focused on a high-$P$ experiment on LSCO to determine the relationship between $\rho_{c}(T)$ and $T_{\rm c}$ regardless of the influence of the structural transition.

According to uniaxial $P$\cite{M.Braden_PRB_93}
and ultrasonic\cite{M.Nahara_PRB_95} measurements on high $T_{\rm c}$ cuprates, the $P$ dependence of $T_{\rm c}$ is characterized by strong anisotropy.
For LSCO, not only the absolute value but also the sign of $dT_{\rm c}/dP$ depends strongly on the direction of the applied $P$ regardless of a change of carrier concentration ($dT_{\rm c}/dP_{\parallel ab} > 0$ and $dT_{\rm c}/dP_{\parallel c} < 0$). \cite{com1} 
Thus, it is necessary to clarify the uniaxial $P$ dependence of $T_{\rm c}$ because the effect of hydrostatic $P$ is given by the sum of the uniaxial $P$ coefficients for each axis 
($dT_{\rm c}/dP = 2\times dT_{\rm c}/dP_{\parallel ab}+dT_{\rm c}/dP_{\parallel c}$).

Our $P$-experiments were performed by using a cubic-anvil device 
\cite{N.Mori_PPAREC_93} with a mixture of Fluorinert FC70 and FC77 as $P$-transmitting medium. 
The samples were put into a cylindrical Teflon cell with an inner space of 1.5 mm diameter and 1.5 mm length, and the current was applied parallel to the cylindrical axis of the cell.  
Vitrification of the fluid medium at low $T$ often causes a slight deviation from hydrostaticity, 
though the applied $P$ is completely hydrostatic while the $P$-medium remains fluid. 
Pressure applied to the anvil unit was held constant within 3\% 
during $T$ sweeps. 
In this system, quasi-hydrostatic $P$ can be generated by isotropic movement of six anvil tops even after the fluid medium vitrifies at low $T$ and high $P$. 
The key feature of our measurement is that, in our cubic anvil device, the hydrostaticity strongly depends on the sample shape and size because of a difference in the compressibility between the sample and the vitrified mixture.

Non-hydrostatic $P$ is generally a hindrance in understanding the $P$ dependence of $T_{\rm c}$.
However, if we can determine the anisotropy of the applied $P$, it is possible to determine the intrinsic behavior of $T_{\rm c}$ even when $P$ is high enough suppress the structural change. 
In spite of the deviation from hydrostaticity, $P$ dependence of $T_{\rm c}$ is given by 
\begin{equation}
T_{\rm c}(P) = T_{\rm c}(0)+ 2 P_{\parallel ab} \frac{dT_{\rm c}}{dP_{\parallel ab}} + P_{\parallel c} \frac{dT_{\rm c}}{dP_{\parallel c}}
\end{equation}
Moreover, the uniaxial $P$ derivatives have been reported for the zero $P$ limit. \cite{M.Nahara_PRB_95}
Therefore, we can determine the anisotropy in the $P$ acting on the sample by comparing the observed $T_{\rm c}(P)$ with $dT_{\rm c}/dP_{\parallel ab}$ and $dT_{\rm c}/dP_{\parallel c}$.

Single crystals of LSCO were grown by a traveling-solvent floating-zone method. 
The Sr concentrations used in this work are $x= 0.1$ and $0.15$, 
as determined by electron-probe micro-analysis. 
Out-of-plane resistivity was measured by a four-probe method under $P$ up to 8 GPa for a "stick" shaped sample with dimensions of $L_{110} \times L_{110} \times L_{001} = 0.2 \times 0.25 \times 0.73$ mm$^3$($x$=0.1) and "plate" shaped samples with dimensions $0.4 \times 0.4 \times 0.06$ mm$^3$ ($x$=0.1) and $0.45 \times 0.45 \times 0.06$ mm$^3$($x$=0.15). 
The electrodes were made of gold paste with a post-heat-treatment. 
We verified that the applied $P$ did not damage the sample by checking that the $\rho_{c}(T)$ curves under the ambient $P$ before and after the $P$ measurement agreed well with each other.

For stick shaped LSCO with $x$=0.1, we measured the $P$ and $T$ dependence of $\rho_{c}$.
Some interesting features were found in the $P$ dependence of the $\rho_{c}(T)$ curve in the vicinity of $T_{\rm c}$ as is shown in Fig. 1, while $\rho_{c}$ at 300K monotonically decreases with $P$ as shown in the inset.
By applying $P \leq 3$ GPa, the maximum value of $\rho_{c}(T)$ just above $T_{\rm c}$ is drastically reduced, though $T_{\rm c}$ changes weakly.
On the other hand, by applying $P \geq 3$ GPa the maximum value of $\rho_{c}(T)$ is almost constant though $T_{\rm c}$ is strongly suppressed.
The feature at around 3GPa is probably due to the structural change from the orthorhombic to the tetragonal phase. 
According to the reported $P$ measurements,
\cite {S.Nakayama_PC_94,H.J.Kim_PC_88} 
the observed $T_{\rm d}\sim 270$K can be suppressed by $P \geq 3$ GPa.
Indeed, a change in the slope of the $\rho_{c}(T)$ curve, which is clear at around $T_{\rm d}\sim 270$K under ambient $P$, is completely suppressed by applying $P \geq 3$ GPa.
Therefore, we focus on measurements of $\rho_{c}(T)$ under $P\geq 3$GPa in order to address the intrinsic variation of $\rho_{c}$ and $T_{\rm c}$ independent of the structural transitions. 
Incidentally, we note that a deviation from hydrostaticity is not so large in this measurement judging from the observed $P$ dependence of $T_{\rm d}$.

For the plate shaped LSCO sample with $x=0.1$ and 0.15, the $P$ and $T$ dependence of $\rho_{c}$ was also measured. 
$P$ dependence of $\rho_{c}(T)$ for the plate sample with $x$=0.15 is representatively shown in Fig. 2.
The $P$ dependence for the plate sample is different from that for the stick sample. 
In particular, increasing $\rho_{c}$ is induced by applying $P \geq 0.8$ GPa at 297K as shown in the inset.
Moreover, the $P$ dependence of the interlayer distance is sensitively reflected in $\rho_{c}$ in LSCO;\cite{F.Nakamura_PRB_96,F.Nakamura_PC_in} 
namely, $\rho_{c}(P)$ can be regarded as a strain gauge along the $c$ axis.
Therefore, we deduce that the $c$ axis is stretched by applying $P$ in this measurement. 
These results obviously indicate that the stress on the plate sample is strongly anisotropic.

We can infer the reason why the stress on the plate sample is strongly anisotropic even though the sample is not in direct contact with the Teflon cell. 
It is empirically known that the Fluorinert mixture is jelled by applying $P \geq 1$ GPa at around 300K. 
The jelled mixture, which has a much larger compressibility than the sample, acts as a cushion. 
Hence, quasi-hydrostatic conditions would be achieved if the sample were embedded in a large volume of mixture. 
However, a deviation from hydrostatic $P$ is likely caused by the strongly anisotropic distribution of the jelled mixture, though the displacement of six anvil-tops was isotropic. 
In the $P$ cell, the $c$ axis of the plate sample was arranged parallel to the cylindrical axis of the cell. 
Comparing the sample size with the inner diameter of the cell, we can see that the clearance between the inner wall of the cell and the sample is much narrower than the space along the $c$-axis direction. 
Pressure transmission is strongly anisotropic because the jelled mixture shows poor fluidity and the cell geometry is anisotropic. 
In the plate sample measurement, the in-plane compression must be much stronger than the interlayer one; 
thus, the $c$-axis stretch is caused by a large contribution from Poisson's ratio.

Figure 3 provides a plot of $T_{\rm c}$ determined by zero resistivity against $P$ 
for the stick shaped sample ($x=0.1$) and plate ones ($x$=0.1 and 0.15). 
Both the stick and plate sample with $x=0.1$ show a change of slopes in $T_{\rm c}(P)$ at around $P_d \sim 3$ GPa. 
Then, $T_{\rm c}$ changes linearly with $P$ in the orthorhombic phase. 
We found a significant difference between $P$-dependence of $T_{\rm c}$ for the stick and plate samples. 
For the stick sample, $T_{\rm c}$ decreases at a rate of $dT_{\rm c}/dP \sim -0.8$K/GPa for $P \leq 3$ GPa, and $T_{\rm c}$ for $P \geq 3$ GPa is more strongly suppressed with a rate of $dT_{\rm c}/dP \sim -3$K/GPa. 
By contrast, $T_{\rm c}$ for the plate sample increases with a rate of $dT_{\rm c}/dP \sim +3$ K/GPa for $P \leq 3$ GPa and $dT_{\rm c}/dP \sim +0.6$ K/GPa under $P \geq 3$ GPa.
The plate sample with $x=0.15$ shows $T_{\rm c}$ enhancement from 38 K at ambient $P$ to 51.6 K at 8 GPa with a rate of $dT_{\rm c}/dP \sim +5$ K/GPa under $P \leq 1.5$ GPa and $dT_{\rm c}/dP \sim +1.5$ K/GPa for $P \geq 1.5$ GPa. 
The maximum $T_{\rm c}=$ 51.6K is the highest recorded so far for LSCO.

We have attempted to quantify the non-hydrostaticity at low $T$ in order to clarify these results. 
It is reported that the uniaxial-$P$ derivatives obtained at around ambient $P$ are $dT_{\rm c}/dP_{\parallel ab} = +3.2$ K/GPa and $dT_{\rm c}/dP_{\parallel c} = -6.6$ K/GPa for $x=0.1$.\cite{M.Nahara_PRB_95} 
Comparing these uniaxial-$P$ derivatives with the observed $dT_{\rm c}/dP$ in the orthorhombic phase, 
we estimated the anisotropy of applied $P$ on the stick and plate samples using Eq. (1). 
In the measurement for the stick sample with $x$=0.1, we obtained the anisotropic $P$ ratio to be $P_{\parallel c}/P_{\parallel ab}\sim $1.1.
By contrast, the $P_{\parallel c}/P_{\parallel ab}$ values for the plate samples with $x$=0.1 and 0.15 are about  0.6 and 0.5, respectively. 
Thus, the above-mentioned inference is probably confirmed by this estimation.
Incidentally, the $c$-axis stretch for the plate samples is also indicated by comparing $P_{\parallel c}/P_{\parallel ab}$ with compressibility \cite{G.Oomi_PC_91} and Poisson's ratio, \cite{Poisson} as is indicated by increasing $\rho_{c}$.

Next, we consider the intrinsic $P$ dependence of $T_{\rm c}$ in the tetragonal phase. 
Assuming that the anisotropic $P$ ratio in the orthorhombic phase is retained in the $P$-induced tetragonal one, 
we estimated the uniaxial-$P$ derivatives in the tetragonal phase to be $dT_{\rm c}/dP_{\parallel c} = -8 \pm 1$ K/GPa and $dT_{\rm c}/dP_{\parallel ab}= +3 \pm 0.5$K/GPa from Eq.(1). 
Equally, we obtained the uniaxial-$P$ derivatives for $x=0.15$ to be $dT_{\rm c}/dP_{\parallel c} = -9 \pm 1$ K/GPa and $dT_{\rm c}/dP_{\parallel ab}= +3.5 \pm 0.5$K/GPa utilizing the previously reported $P$ dependence of $T_{\rm c}$.\cite{F.Nakamura_PRB_96}
Though $dT_{\rm c}/dP_{\parallel ab}$ remains nearly constant regardless of the structural change, 
$dT_{\rm c}/dP_{\parallel c}$ in the tetragonal phase is much stronger than that in the orthorhombic phase. 
Incidentally, from a sum of these uniaxial $P$ derivatives we obtained a hydrostatic $P$ derivative in the $P$-induced tetragonal phase to be $dT_{\rm c}/dP \sim -1$K/GPa for optimal doping. 
This estimation agrees well with the reported value of $dT_{\rm c}/dP=-1$K/GPa for polycrystalline LSCO with $x$=0.15 under hydrostatic $P$ up to 8GPa.\cite{N.Mori_PCOS_92}
Therefore, it is a reasonable assumption that the anisotropic $P$ ratio is retained in the high-$P$ range.

What we are mainly interested in now is why the absolute value of $dT_{\rm c}/dP_{\parallel c}$ 
is much stronger than that of $dT_{\rm c}/dP_{\parallel ab}$. 
In the tetragonal phase, the enhancement of $T_{\rm c}$ due to interlayer expansion is about 2.6 times stronger than that due to in-plane compression. 
Naturally, the $dT_{\rm c}/dP_{\parallel ab}$ value is too small to account for the observed $P$ dependence of $T_{\rm c}$.  
As shown in Fig. 4, the variations of $T_{\rm c}$ obtained for both stick and plate shaped samples could be scaled linearly with $\rho_{c}$ at 297K. 
Additionally, the $P$ dependence of $\rho_{c}$ can be expressed as a function of the lattice parameter $c$. \cite{F.Nakamura_PRB_96,F.Nakamura_PC_in}
Therefore, $T_{\rm c}$ is much more strongly reflected in the interlayer distance than the in-plane one. 
Thus, a change of $T_{\rm c}$ is mainly interpreted 
in terms of the interlayer coupling as monitored in $\rho_{c}$.

Incidentally, $\rho_{c}$ is given by $\rho_{c} \sim [ N |t_c|^2 \tau ]^{-1}$, 
where $N$ is carrier number, 
and $\tau$ is lifetime in the plane. \cite{N.P.Ong_PC_94}
In general, $P$ dependence of $N$ and $\tau$ is reflected 
in those of the residual resistivity and the slope of $\rho_{ab}(T)$, respectively.
\cite{T.M.Rice_PRB_72}
Recently, we have reported that anisotropic pressure does not change $\rho_{ab}$ below about 60 K. \cite{Goko_JPS,F.Nakamura_PC_in} 
Moreover, weak $P$ dependence of $N$ has been reported. \cite{C.Murayama_PC_91,G.Q.Zheng_PC_95}
Therefore, we deduce that $\tau$ and $N$ for LSCO are almost independent of $P$ at low $T$; thus $P$ dependence of $\rho_{c}$ for LSCO is mainly governed by that of $t_c$.

As a clue to understanding the relation between $T_{\rm c}$ and interlayer coupling $t_c$, we focus on the low-$T$ behavior of $\rho_{c}$ in the tetragonal phase when the $c$-axis compression is stronger than the in-plane one. 
Although the $P$ reduces $\rho_{c}$ values over the whole $T$ range, the peak value of $\rho_{c}(T)$ just above $T_{\rm c}$ is almost independent of $P$ as shown in Fig. 1. 
This observation suggests that superconductivity occurs when $t_c$ reaches some critical value.
Thus, higher $T_{\rm c}$ values can be expected in the system which has strongly suppressed $t_c$.

We infer why a strongly 2D electronic state is an advantageous factor for the superconductivity in LSCO against the expectations from the theoretical model. \cite{J.M.Wheatly_PRB_88} 
$T$ dependence of $\rho_{c}$ and $\rho_{ab}$ indicate that 
the interlayer coupling is suppressed with decreasing $T$. 
It seems that the perfectly 2D metal expected from the spin-charge separation model is achieved at absolute zero Kelvin when superconductivity is absent.
However, it is believed that the perfectly 2D metal is unstable at low $T$. 
Moreover, it has been reported that not only $\rho_{c}$ but also $\rho_{ab}$ shows a logarithmic upturn when $T_{\rm c}$ is suppressed by a large magnetic field. \cite{Y.Ando_PRL_95}
Therefore, superconductivity or a 3D localized electronic state is likely required at low $T$ in order to depress the instability. 
Thus, enhancing the instability, which is caused by the $c$-axis expansion, increases $T_{\rm c}$.

In conclusion, we have demonstrated that strong 2D structure of the electronic state 
as is monitored in $\rho_{c}$ is a key parameter for high-$T_{\rm c}$ superconductivity, 
though low dimensionality is generally known as a destructive factor for conventional superconductivity. 
The remarkable thing is that $T_{\rm c}$ is strongly reflected in interlayer distance 
rather than in-plane one.  
Indeed, the enhancement of $T_{\rm c}$ in LSCO, which reaches 51.6 K at 8 GPa, 
is mainly interpreted in terms of interlayer expansion, which enhances two dimensionality. 
Therefore, we expect that measurement under anisotropy controlled pressures 
causes much higher $T_{\rm c}$ than previously reported values for many high-$T_{\rm c}$ materials.


%
We would like to thank S. R. Julian for criticizing the manuscript. This work was supported by a Grant-in-Aid for Scientific Research on Priority Areas, 
from the Ministry of Education, Science, Sports and Culture of Japan and a grant from NEDO. 
%
%
%

%
%
\begin{figure}
\begin{center}
  \epsfxsize=8cm
   \epsfbox{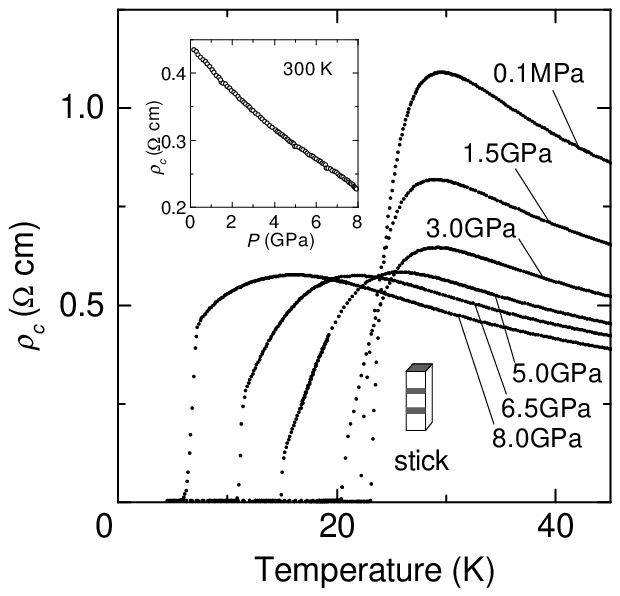}
   \caption{ $\rho_{\rm c}(T)$ curves measured for stick shaped 
LSCO sample with $x$=0.1 below 55K under $P$ up to 8 GPa. 
The inset shows $P$ dependence of $\rho_{c}$ at 300K. The sample shape and the configurations of the electrodes are shown. }
   \label{LB7253Fig1.eps}
\end{center}
\end{figure}
%
%
%
%
\begin{figure}
\begin{center}
  \epsfxsize=8cm
   \epsfbox{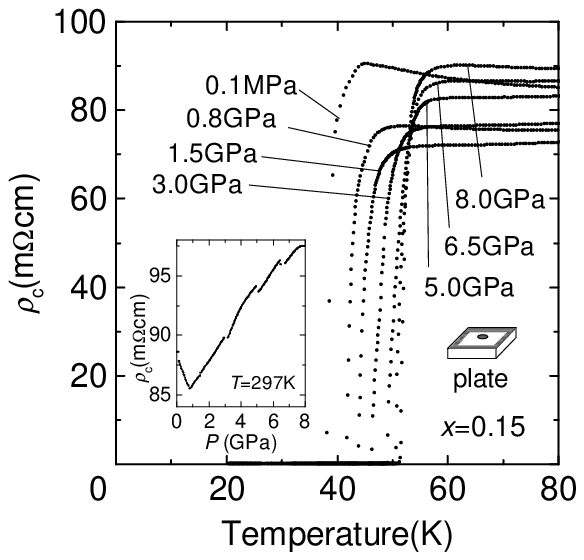}
   \caption{ $\rho_{c}(T)$ curves measured for the plate shaped 
LSCO sample with $x$=0.15 under $P$ up to 8 GPa. The inset shows $P$ dependence of $\rho_{c}$ at 297K. 
Also shown is a schematic representation of the electrodes attached to the plate shaped crystals.}
\label{LB7253Fig2.eps}
\end{center}
\end{figure}
%
%
%
%
\begin{figure}
\begin{center}
  \epsfxsize=8cm
     \epsfbox{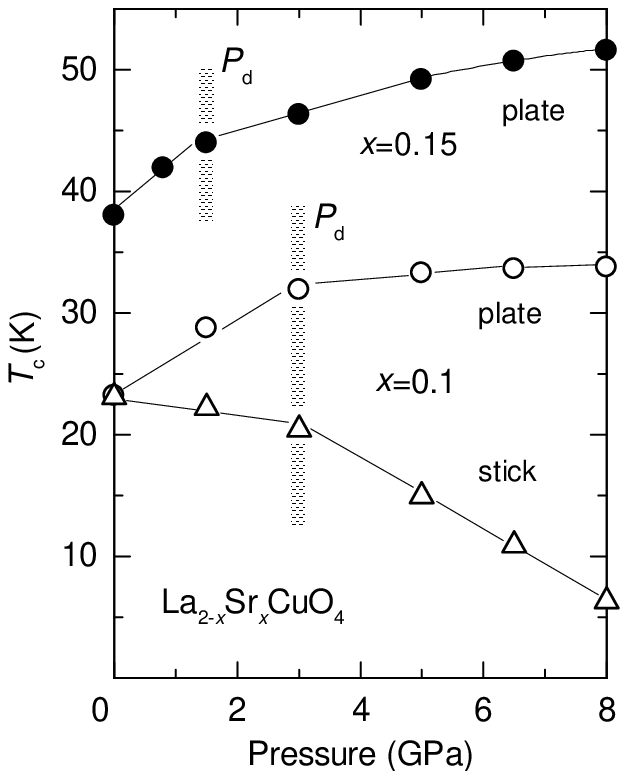}
    \caption{ $T_{\rm c}$ for stick and plate shaped samples against $P$. 
Solid line is a guide to the eye.
             Meshed marks indicate the structural transition pressure 
             $P_{\rm d}$ estimated from the changes in $\rho_{c}(T)$ 
             and the reported $T_{\rm d}(P)$
             (Ref. 14,15).}
\label{LB7253Fig3.eps}
\end{center}
\end{figure}
%
%
\begin{figure}
\begin{center}
  \epsfxsize=8cm
   \epsfbox{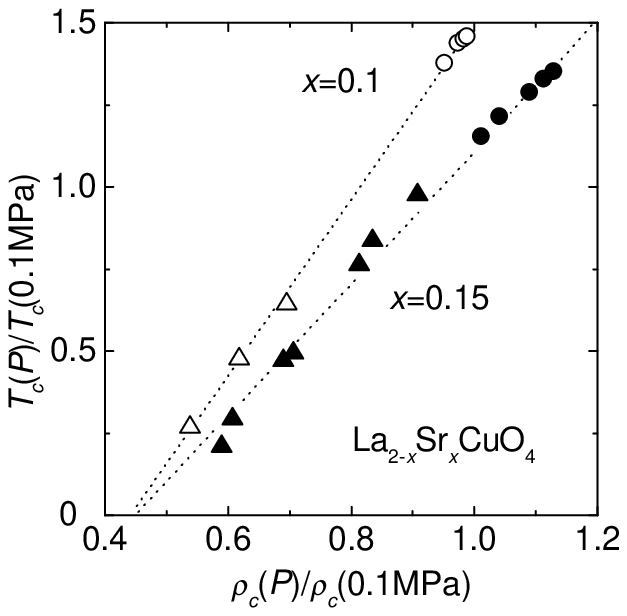}
\caption{ The variation of $T_{\rm c}(P)/T_{\rm c}$(0.1MPa) in the tetragonal phase plotted as a function of 
$\rho_{c}(P)/\rho_{c}$(0.1MPa) at 297K.
Circles and triangles indicate $T_{\rm c}$ for plate and stick shaped samples, respectively. 
$T_{\rm c}$ for the stick LSCO with optimal doping (closed triangle) is obtained from Ref. [7,13].}
\label{LB7253Fig4.eps}
\end{center}
\end{figure}
%
%
%
%

\begin{references}
\bibitem{com1}
We use the value $dT_{\rm c}/dP_{\parallel ab}$ which is the average of $dT_{\rm c}/dP_{\parallel a}$ 
and $dT_{\rm c}/dP_{\parallel b}$ because LSCO spontaneously forms a twinned structure below $T_{\rm d}$.
\bibitem{Poisson}
We use a Poissons ratio $\sim 0.3$ 
obtained from our uniaxial-$P$ experiment 
for LSCO with $x = 0.15$.
\end{references}
\end{document}